\documentclass[journal=jpccck,manuscript=article]{achemso}
\usepackage{amsmath}

\newcommand*{\beq}[0]{\begin{equation}}
\newcommand*{\eeq}[0]{\end{equation}}
\newcommand*{\beqs}[0]{\begin{eqnarray}}
\newcommand*{\eeqs}[0]{\end{eqnarray}}
\newcommand*{\vasp}[0]{\textsc{vasp}}

\newcommand*{\Eqn}[1]{\ref{eqn:#1}}
\newcommand*{\Fig}[1]{\ref{fig:#1}}
\newcommand*{\Tab}[1]{\ref{tab:#1}}

\newcommand*{\TiO}[0]{TiO$_2$}
\newcommand*{\TiOr}[0]{TiO}
\newcommand*{\x}[0]{\times}
\newcommand*{\et}[0]{\textit{et al.}}
\newcommand*{\kB}[0]{k_{\text{B}}}

\author{Min Yu}
\affiliation{Department of Physics, University of Illinois at
Urbana-Champaign, Urbana, IL 61801}
\author{Dallas R. Trinkle}
\email{dtrinkle@illinois.edu}
\affiliation{Department of Materials Science and Engineering, University of
Illinois at Urbana-Champaign, Urbana, IL 61801}

\title{Au/\TiO(110) interfacial reconstruction stability from ab initio}

\keywords{Interfacial structure, Nanoparticle, Titania}

\begin{document}

\begin{abstract}
We determine the stability and properties of interfaces of low-index Au surfaces adhered to \TiO(110), using density functional theory energy density calculations.  We consider Au(100) and Au(111) epitaxies on rutile \TiO(110) surface, as observed in experiments.  For each epitaxy, we consider several different interfaces: Au(111)//\TiO(110) and Au(100)//\TiO(110), with and without bridging oxygen, Au(111) on $1\x2$ added-row \TiO(110) reconstruction, and Au(111) on a proposed $1\x2$ \TiOr\ reconstruction.  The density functional theory energy density method computes the energy changes on each of the atoms while forming the interface, and evaluates the work of adhesion to determine the equilibrium interfacial structure.
\end{abstract}

\section{Introduction}
Bulk metallic Au is chemically inert and catalytically inactive as a consequence of combination of valence $d$ orbitals and diffused valence $s$ and $p$ orbitals. Recently, Au nanoparticles have been found to be catalytically active when supported on metal oxides such as \TiO, SiO$_2$, Fe$_2$O$_3$, Co$_3$O$_4$, NiO, Al$_2$O$_3$, MgO, etc.\cite{Au-TiO2:Haruta1987,Au-TiO2:Haruta1989,Au-TiO2:Valden1998,Au-TiO2:Boccuzzi1989,Au-TiO2:Hayashi1998,TiO2:SSReport} For example, Au nanoparticles supported on a \TiO(110) surface demonstrate catalytic activity to promote the reaction between CO and O$_2$ to form CO$_2$ at $T < 40\,$K with 3.5\,nm Au nanoparticles maximizing activity.\cite{Au-TiO2:Valden1998} The catalytic activity is remarkably sensitive to the support material, Au particle size and Au-support interaction; in addition, the reaction mechanism of CO oxidation over Au/TiO$_2$ system remains under debate.\cite{Au-TiO2:Valden1998,Au-TiO2:Bond2000,Au-TiO2:Grunwaldt1999,Au-TiO2:Haruta2004} High-resolution transmission electron microscopy (HRTEM)\cite{Au-TiO2:Cosandey2001,Au-TiO2:Shankar} and high-angle annular dark field scanning transmission electron microscopy (HAADF-STEM)\cite{Au-TiO2:Shankar,Au-TiO2:Akita2008} have characterized the atomic structure of nanocrystal interface.  However, the atomic structure of Au/\TiO\ interface is difficult to determine in HRTEM image simulations due to several issues, such as the thickness of nanoparticles and metal oxide substrates are not determined, and the positions of atoms in the direction parallel to the electron beam are not determined, and the very low contrast for oxygen atoms.  New HRTEM experiments\cite{Au-TiO2:Shankar} observed Au nanoparticles on \TiO (110) surfaces with both the Au(111) and Au(100) epitaxies, with the Au(111) epitaxy more frequently observed than Au(100). Their analysis with HAADF-STEM analyzed the reconstructed interface of epitaxial Au(111) sitting on a \TiO (110) $1\x2$ surface, and extracted important geometric information such as interlayer separations, the presence of Au in the interface of a $1\x2$ reconstruction, and estimates of the work of adhesion.

Density functional theory (DFT) calculations\cite{Theor:KS} have studied the optimum size and stable adsorption of Au nanoparticles on rutile \TiO(110).  Single Au atom is energetically favorable on the atop site above five-fold coordinated (5c) Ti atom on a stoichiometric \TiO\ surface,\cite{Au-TiO2:Yang2000} and is most stable on the two-fold coordinated (2c) bridging O vacancy site on a reduced surface.\cite{Au-TiO2:Wang2003,Au-TiO2:Amrendra2003,Au-TiO2:Wahlstrom2003}  Oxygen vacancies cause a stronger binding of Au atoms,\cite{Au-TiO2:Okazaki2004} nanoclusters\cite{Au-TiO2:Lopez2004,Au-TiO2:Pabisiak2009,Au-TiO2:Pillay2005} and nanorows\cite{Au-TiO2:Pabisiak2009} to the reduced \TiO\ surface than to the stoichiometric surface.  Apart from the stoichiometric and reduced \TiO\ surfaces, Shi \et\ found the O-rich interface is the most stable at low temperature of catalytic reaction after examining the Au-rod/\TiO (110) in the orientation Au(111)//\TiO(110) with different interface stoichiometry and various rigid-body translations.\cite{Au-TiO2:Shi2009}  Recently, Shibata~\et\ examined two and nine Au(110) atomic layers supported on reduced \TiO(110), and demonstrated that both the atomic and electronic structure of two-layer Au are reconstructed, while the lattice coherency decays rapidly across the interface for nine-layer Au.\cite{Au-TiO2:Shibata2009} We compare different Au/\TiO\ interfaces: Au(111)//\TiO(110) and Au(100)//\TiO(110), with and without bridging oxygen, Au(111) on $1\x2$ added-row \TiO(110) reconstruction\cite{TiO2:AddedRow}, and Au(111) on a new proposed $1\x2$ \TiOr\ reconstruction\cite{Au-TiO2:Shankar}.  We use the newly-reformulated\cite{EDM:ChettyMartin,EDM:PAW} density functional theory energy density method to evaluate energy for each atom in the interfacial reconstruction.  This provides insight into interfacial stability from the changes in atomic energy from the formed interface, and corrects for spurious errors in the work of adhesion from the remaining free surfaces in the computational cell.  The new information of atomic energies extracted from density functional theory shows the response to bonding environment changes in interfaces.  The comparison with experimental geometry\cite{Au-TiO2:Shankar} and work of adhesion\cite{Au-TiO2:Shankar2010} allows us to validate our predicted structures.

\section{Methodology}
\label{sec:EDM}
We perform DFT calculations\cite{Theor:KS} on the Au/\TiO\ interfaces using the projector augmented wave (PAW) method\cite{Theor:PAW} with the Vienna ab initio simulation package (\vasp)\cite{Theor:VASP,Theor:VASP-USPP-PAW}. The exchange-correlation energy is treated in the Perdew-Burke-Ernzerhof\cite{Theor:PBE} version of the generalized gradient approximation functional (PBE-GGA).  Elements Au, Ti, and O are given by [Xe]$6s^{1}5d^{10}$, [Ne]$3s^{2}3p^{6}4s^{2}3d^{2}$, [He]$2s^{2}2p^{4}$ atomic configurations; this requires a plane-wave basis set cut-off at 900\,eV. We use Monkhorst-Pack k-point meshes\cite{Theor:MP} of $1\x6\x1$ for interface supercells; Brillouin-zone integration uses the Methfessel-Paxton method\cite{Methfessel1989} with $\kB T = 0.2\,$eV for electronic occupancies, and the total energy extrapolated to $\kB T = 0\,$eV. The calculated lattice constant for Au in the FCC phase is 4.171\,\AA, and for \TiO\ in the rutile phase $a=4.649\,$\AA, $c=2.970\,$\AA, and $u=0.305$.  These calculated values compare well with the experimental values of 4.08\,\AA\ for Au and $a=4.584\,$\AA, $c=2.953\,$\AA, $u=0.305$\cite{TiO2:SSReport} for \TiO.

The work of adhesion of forming an interface from two individual surfaces can be determined from total energy calculations:
\beq
E_{\text{adh}}=\frac{1}{A}(E_{\text{Au}}+E_{\text{\TiO}}-E_{\text{Au/\TiO}}),
\label{eqn:totalE}
\eeq
where $E_{\text{Au}}$ and $E_{\text{\TiO}}$ are the energy of relaxed Au surface and relaxed \TiO\ surface and $E_{\text{Au/\TiO}}$ is the energy of the interface system.  To avoid differences in grid densities or the planewave basis, the surfaces are computed with the same supercell as the interface system.  In addition to total energies, the energy density method proposed by Chetty and Martin\cite{EDM:ChettyMartin} provides the formation energy for more than one surface or interface in one calculation, and a picture of the distribution of energy among the surrounding atoms.  We use a new reformulation of the energy density method with \vasp\ for PAW method.\cite{EDM:PAW}  Moreover, we compute atomic energies by integrating the local energy density over gauge independent integration volumes.\cite{WeightMethod}  The data allows us to identify the spatial range of the interface and gives insight into the nature of interfacial stability.  The integration of the energy density over these volumes produces a small integration error, that can be estimated from the extent to which gauge-invariance is broken; we include that error as a $\pm$ range in all of our reported energy density calculations.  For the Au/\TiO\ interfaces, the supercell configurations in the calculations are periodic parallel to the interface, and contain six layers of Au, eight trilayers of \TiO, and 10.5\,\AA\ vacuum region.  Due to the lattice mismatch, Au layers are strained to lattice match the \TiO\ according to the supercell periodicity; strained Au surfaces are used as references for energy differences.  Atomic relaxation is allowed for all six layer Au atoms and for three interfacial layers of \TiO\ for all geometries considered.  In addition, different translations of Au relative to \TiO\ are attempted in order to determine the minimum energy configuration.  The equilibrium positions of the atoms are determined by requiring the force on each relaxed atom to be smaller than 0.02\,eV/\AA.

\section{Interfaces}

\begin{figure}[htp]
 \includegraphics[width=3in]{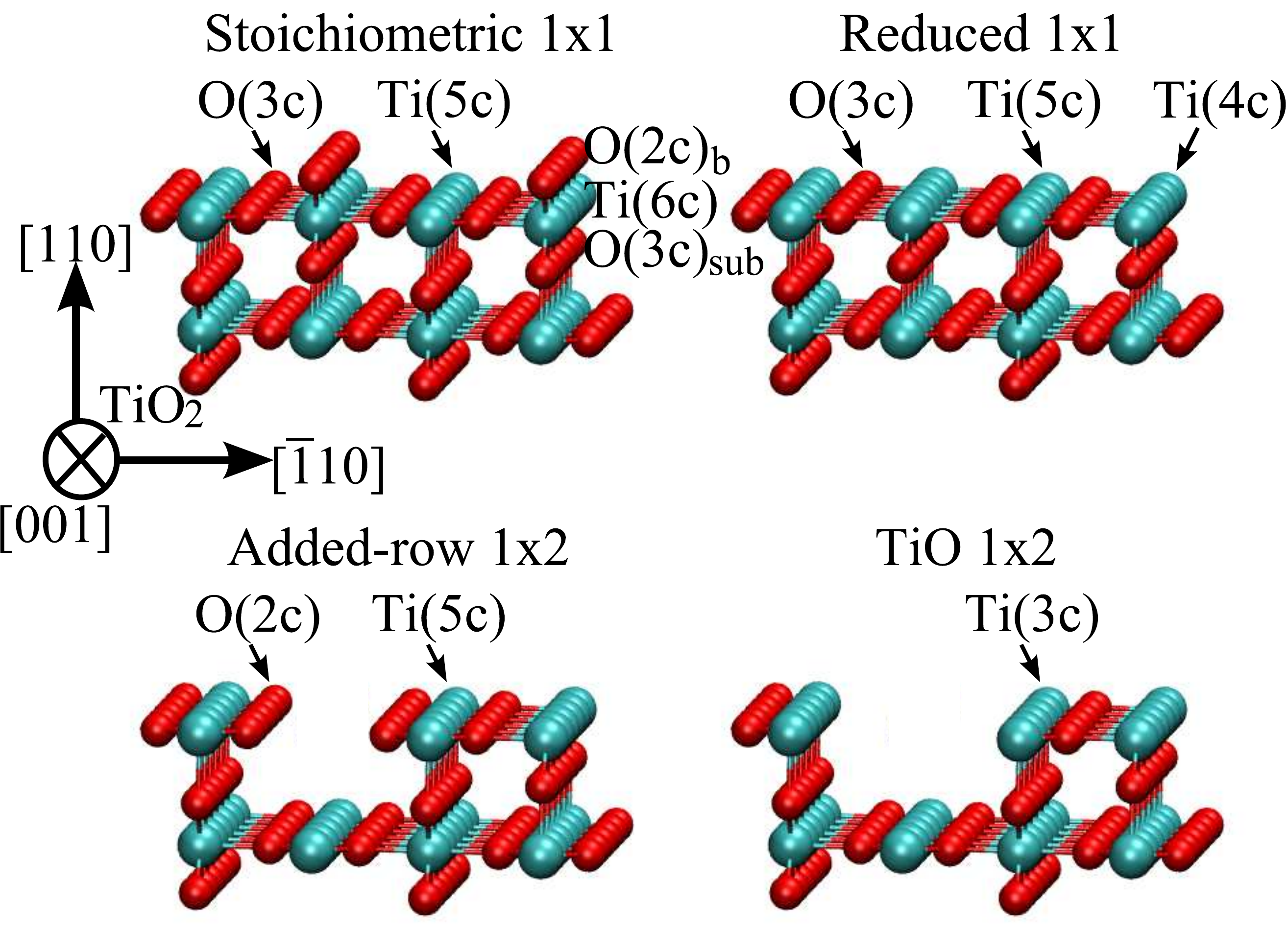}
\caption{Geometry for four different \TiO (110) surface structures.  Upper two are stoichiometric $1\x1$ and reduced $1\x1$; bottom two are added-row $1\x2$ reconstruction and \TiOr\ $1\x2$ reconstruction.  The stoichiometric structure has bridging oxygen (2c)$_\text{b}$ atoms above the flat titanium (5c) / (6c)  and oxygen (3c) plane.  Removal of the bridging oxygens produces a reduced surface, with four- and five-fold coordinated titanium.  The added-row reconstruction removes every other row of Ti (4c) atoms with subsurface bridging oxygens (3c)$_\text{sub}$ for a $1\x2$ reconstruction, with two-fold coordinated oxygen.  Finally, additional reduction of the added-row reconstruction, by removing the oxygen (2c) atoms neighboring the removed row, produces the \TiOr\ reconstruction with three-fold coordinated titanium.}
\label{fig:TiO2-110-2L}
\end{figure}

\Fig{TiO2-110-2L} shows the four different configurations of rutile \TiO (110) substrates we consider.  We start with a stoichiometric surface, and then reduce the surface by removing all bridging O atoms; both are $1\x1$ surfaces.  Pang~\et\ proposed an added-row $1\x2$ reconstruction for the rutile (110) surface, where one row of Ti atom with its sub-bridging O row are removed per $1\x2$ cell for a fully reduced surface.\cite{TiO2:AddedRow}  Finally, removing the two-fold coordinated O atoms from the added-row reconstruction gives a \TiOr\ reconstruction.  While this reconstruction is not the lowest in energy, it provides the most stable Au/\TiO\ interface that also matches the experimentally observed geometry.

\subsection{Au(111)//\TiO(110) $1\x 1$: Stoichiometric and reduced interfaces}

Both interfaces on $1\x 1$ surfaces use a similar geometry for relaxation.  Along the direction Au$[1\bar10]$//\TiO[001], a single repeat length of Au and \TiO\ gives a 1\%\ lattice mismatch.  This agrees with STEM measurements showing registry even up to 10 layers from the interface.\cite{Au-TiO2:Shankar}  Along the direction Au$[\bar1\bar12]$//\TiO$[\bar110]$, a repeat length of 4 for Au matches with a repeat length of 3 for \TiO, producing a total $3.6\%$ lattice mismatch strain at the interface.  The supercells contain 48 Au, 48 Ti, and 96 O atoms in the interface configuration with stoichiometric \TiO\ surface, and 3 fewer O atoms for the reduced \TiO\ surface.  After relaxation, we determine the interlayer spacing at the interface; with energy density calculations, we can ignore any spurious energy changes due to the opposing Au and \TiO\ surfaces.

\begin{figure}[htp]
 \includegraphics[width=3in]{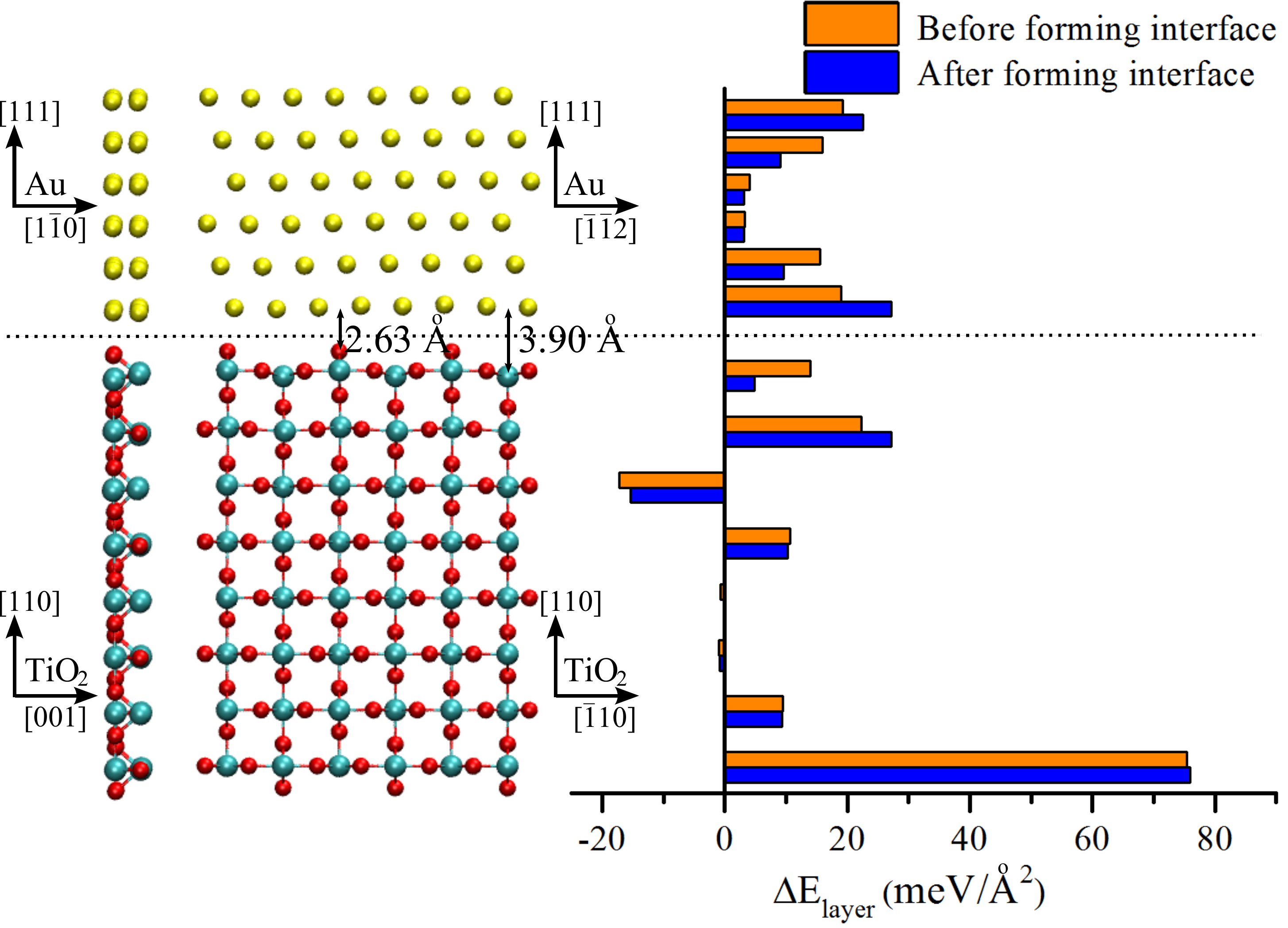}
\caption{Geometry and energy of Au(111) on the stoichiometric \TiO (110) surface following relaxation.  The atomic energy on each layer is referenced to the bulk, and shown before (orange) and after (blue) forming the interface.  The interfacial distance relaxes to 3.90\,\AA\ between Au and Ti layers, and 2.63\,\AA\ between Au and bridging-O layers.}
\label{fig:111+bO}
\end{figure}

\begin{figure}[htp]
 \includegraphics[width=3in]{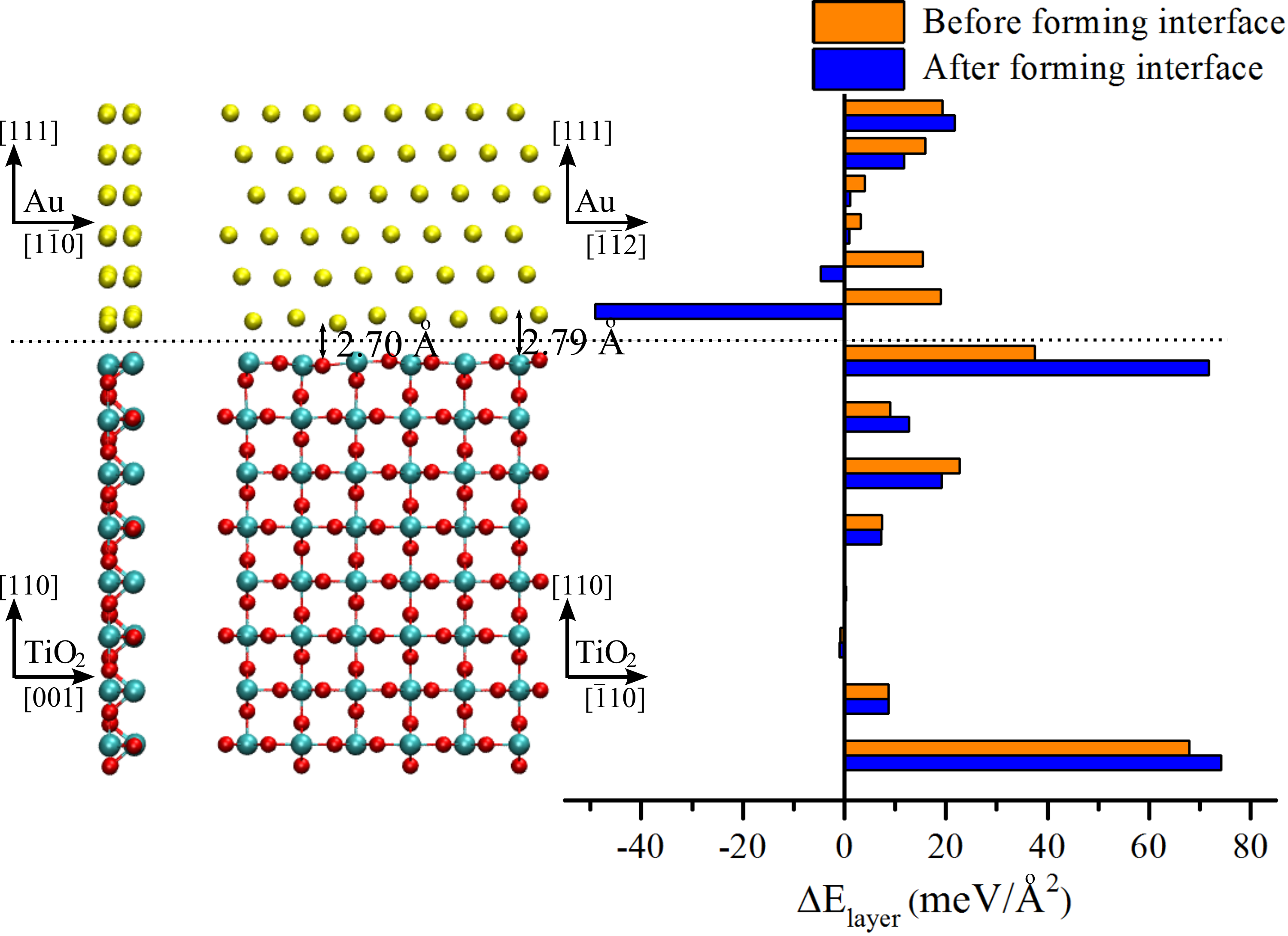}
\caption{Geometry and energy of Au(111) on the reduced \TiO (110) surface following relaxation.  Energy per layer in the reference of bulk value is given before (orange) and after (blue) forming the interface.  The interfacial distance relaxes to 2.79\,\AA\ between Au and Ti layers; and 2.70\,\AA\ between Au and in-plane O layers.  The geometry reduces the energy of the surface Au layer to a more stable configuration than the stoichiometric \TiO\ surface.}
\label{fig:111-bO}
\end{figure}

\Fig{111+bO} and \Fig{111-bO} show the geometry of the relaxed Au(111) on stoichiometric and reduced \TiO (110) surfaces.  The interfacial distance between Au and Ti layers relaxed to 3.90\,\AA\ with stoichiometric \TiO\ surface, and 2.79\,\AA\ in the configuration with reduced \TiO\ surface.  From total energy, the work of adhesion of the interface with stoichiometric \TiO\ surface is 7\,meV/\AA$^2$, while the work of adhesion of the interface with the reduced \TiO\ surface is 54\,meV/\AA$^2$.  The differences in interlayer spacing and energy is due to the presence or absence of bridging oxygen atoms on the \TiO\ surface.  Energy density shows that \TiO\ layers reach bulk behavior by the fifth layer from the interface.  We integrate the energy density over two Au layers and four \TiO\  layers to evaluate the work of adhesion strictly from changes in energy near the interface.  This gives a work of adhesion of $4\pm 2\,$meV/\AA$^2$ to the stoichiometric \TiO\ surface, and $53\pm1\,$meV/\AA$^2$ to the reduced \TiO\ surface.  The work of adhesion is primarily due to a decrease in energy of the Au surface layer at the reduced \TiO\ surface.  This suggests that the main effect of removing bridging oxygen is to provide a flat surface for Au(111) layers to adhere, and that the \TiO\ surface energy change is significantly less than the Au surface energy change.

\subsection{Au(111)//\TiO(110) $1\x2$: Added-row and \TiOr\ reconstructions}

Both interfacial reconstructions on $1\x2$ surfaces use a similar geometry for relaxation.  Along the direction Au$[1\bar10]$//\TiO[001], a single repeat length of Au and \TiO\ gives a 1\%\ lattice mismatch as for the $1\x1$ reconstructions.  Along the direction Au$[\bar1\bar12]$//\TiO$[\bar110]$, a repeat length of 5 for Au matches with a repeat length of 4 for \TiO, producing a total $2.9\%$ lattice mismatch strain at the interface; the different periodicity is required for a $1\x2$ reconstruction. The supercells contain 62 Au, 62 Ti, and 122 O atoms in the interface configuration with added-row \TiO\ reconstruction, and 4 fewer O atoms for the \TiOr\ reconstruction.  After relaxation, we determine the interlayer spacing at the interface; with energy density calculations, we can ignore any spurious energy changes due to the opposing Au and \TiO\ surfaces.

\subsubsection{Added-row reconstruction}

\begin{figure}[htp]
 \includegraphics[width=3.0in]{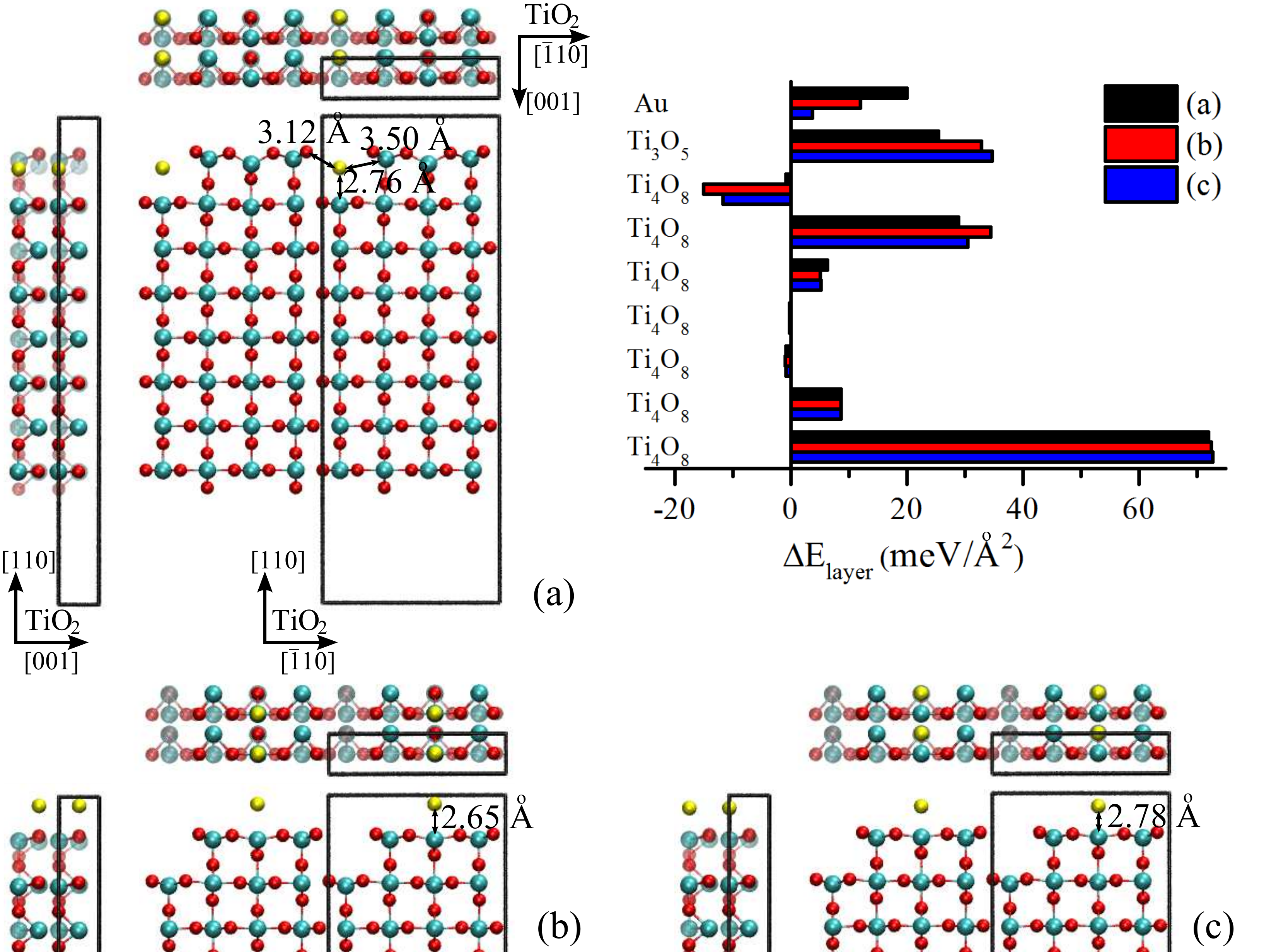}
\caption{Three different configurations of a single Au row on the \TiO\ added-row reconstruction: (a) in the missing Ti row with 4 nearest neighboring O atoms; (b) on top of the \TiO\ surface directly above a Ti atom; (c) on top of the \TiO\ surface bridging between two Ti atoms.  Au atoms are in gold, and the wireframe shows the supercell.  Opaque atoms are on the top layer while transparent atoms are on lower layers.  The (c) configuration has lowest total energy, 6.3\,meV/\AA$^2$ lower than (b) configuration, and 15.7\,meV/\AA$^2$ lower than (a) configuration.   From the energy calculations, the Au row controls the total energy, with the largest increase in energy from filling the missing Ti-O row in the surface layer; hence, we expect to see a mixing of the \TiO\ surface with Au only \textit{after} coverage by a gold nanoparticle.}
\label{fig:1Au+TiO2AddedRow}
\end{figure}

The added-row reconstruction for the $1\x2$ rutile (110) surface removes one row of Ti atom with its sub-bridging O row per $1\x2$ cell for a fully reduced surfaces.\cite{TiO2:AddedRow}  Experimental observations of the interface find a mixed \TiO-Au layer with $1\x2$ periodicity;\cite{Au-TiO2:Shankar} to build our interface and compute the work of adhesion, we consider different configurations to attach a row of Au atoms on added-row reconstruction in \Fig{1Au+TiO2AddedRow}.  After geometry relaxation, the configuration of each Au atom sitting on the top of two Ti atoms with 4 neighboring O atoms is the most stable; there is an energy cost of 15.7\,meV/\AA$^2$ to place a Au row into the missing row of \TiO.  This is similar to the adhesion of Au rows to bridging oxygen vacancies in a \TiO (110) ``missing row'' reconstruction.\cite{Au-TiO2:Pabisiak2009}  The energy density shows that the energy of Au dominates the stability.

\begin{figure}[htp]
 \includegraphics[width=3.0in]{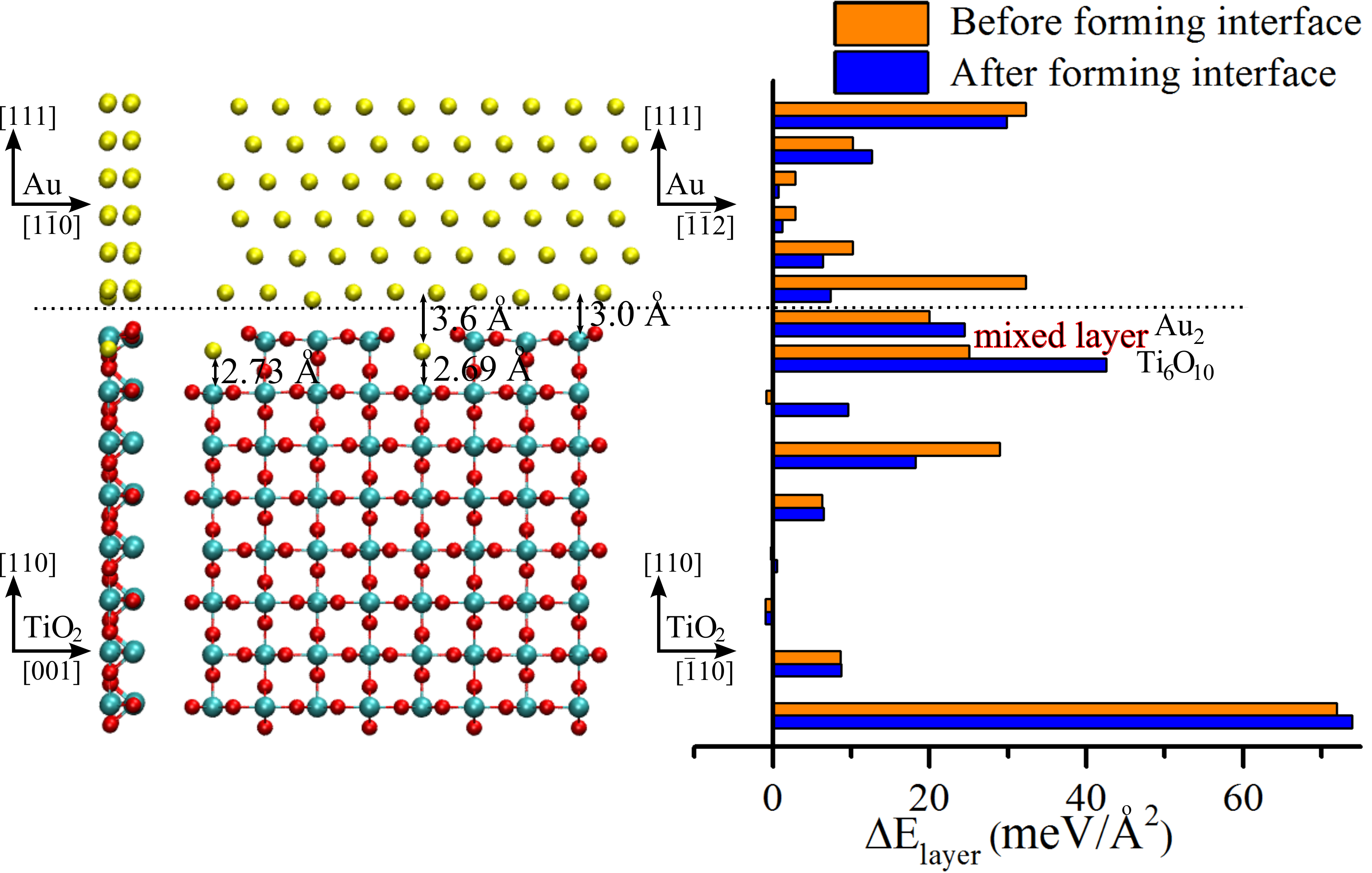}
\caption{Geometry and energy of the Au(111)//\TiO\ added-row reconstruction following relaxation.  Atomic energy per layer in the reference of bulk value is given before (orange) and after (blue) forming the interface.  The interfacial distance between Au layer with mixed layer is about 3.4\,\AA, and the work of adhesion is --9\,meV/\AA$^2$.  While the Au surface layer reduces its energy, the \TiO\ layer increases in energy as the oxygen atoms that neighbor the in-surface Au rows are unable to relax out of the (110) plane; hence, the Ti$_6$O$_{10}$ layer increases in energy.}
\label{fig:AddedRow}
\end{figure}

\Fig{AddedRow} shows the geometry of the relaxed Au(111) on added-row \TiO\ reconstruction.  The interfacial distance between Au and the mixed interfacial layer is 3.4\,\AA.  This larger distance is due to the displacement of oxygen atoms neighboring the interfacial Au rows.  From total energy, the work of adhesion of the interface is --9\,meV/\AA$^2$ after accounting for the 16meV/\AA$^2$ increase in energy due to the addition of Au into the subsurface (c.f. \Fig{1Au+TiO2AddedRow}).  We integrate the energy density over two Au interfacial layers, one mixed interfacial layer and three next \TiO\ interfacial layers and subtract the corresponding energy density integration in Au layers and the ground-state configuration of an Au row on \TiO, \Fig{1Au+TiO2AddedRow}(a).  This energy density calculation gives a work of adhesion of $6\pm1\,$meV/\AA$^2$ before subtracting 16meV/\AA$^2$.  After forming interface, the atomic energy of Au interfacial layer drops, while the atomic energy of \TiO\ in the mixed layer increases.  The increase in the energy of the surface Ti$_6$O$_{10}$ layer is due to the constraint placed on oxygen atoms neighboring to the intermixed Au row in the mixed layer.

\subsubsection{\TiOr\ reconstruction}
 
\begin{figure}[htp]
 \includegraphics[width=3.0in]{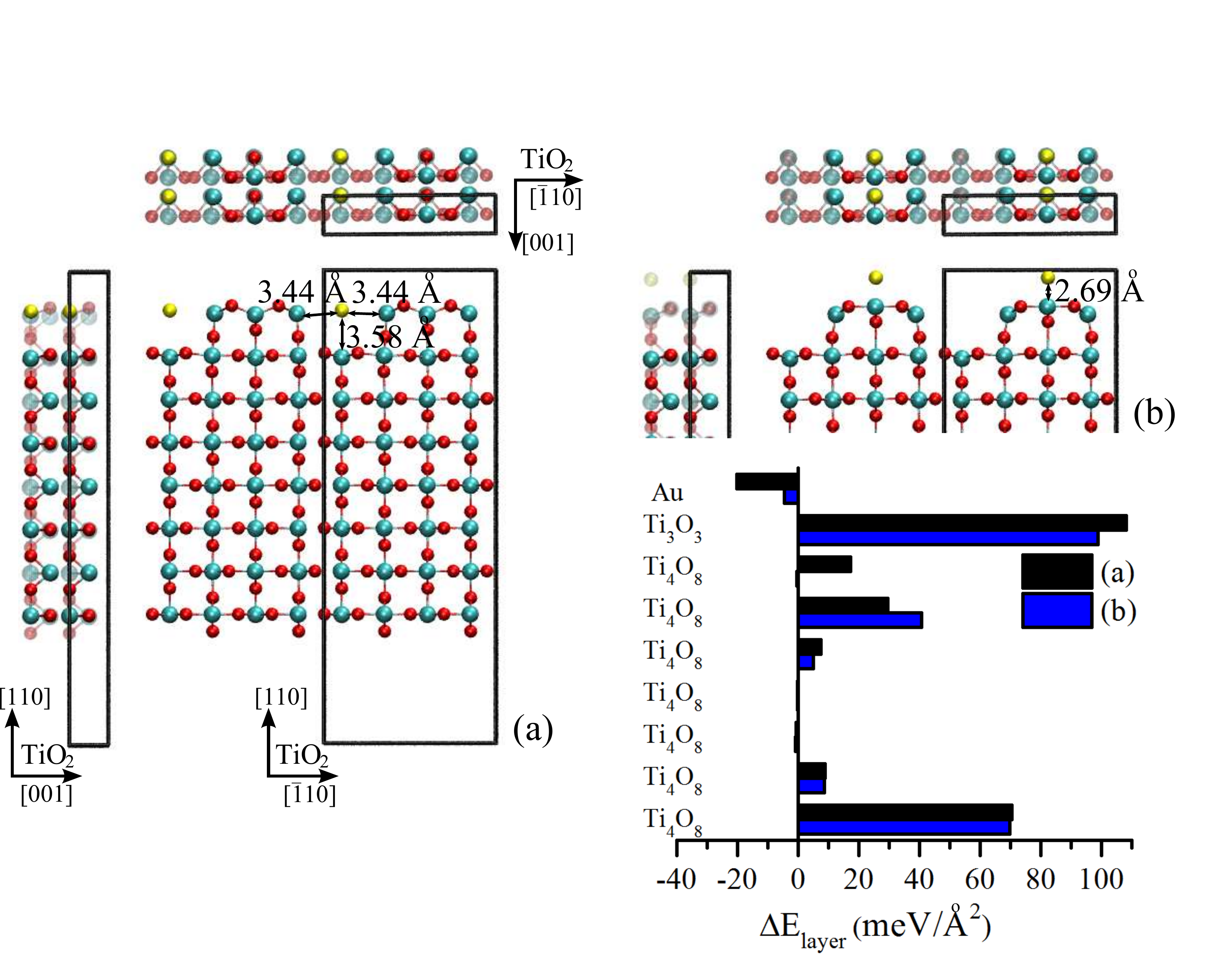}
\caption{Two different configurations of a single Au row on the \TiOr\ reconstruction: (a) in the missing Ti row; (b) on top of the \TiO\ surface bridging between two Ti atoms.  Au atoms are in gold, and the wireframe shows the supercell.  Opaque atoms are on the top layer while transparent atoms are on lower layers.  The energy of the (b) configuration is 0.8\,meV/\AA$^2$ lower than (a) configuration.  Adding Au into the missing row only slightly increases the energy of the Au row; this increase is much less than for the \TiO\ added-row reconstruction.  However, the \TiOr\ reconstruction is a higher energy surface than the added-row reconstruction.}
\label{fig:1Au+TiO}
\end{figure}

The added-row reconstruction can be further reduced by removing the two-fold coordinated O atoms on the \TiO\ surface layer to form a \TiOr\ $1\x2$ reconstruction.  This reconstruction is suggested by the energy density calculations above as a possible route to increase the work of adhesion.  We build our interface in a similar manner as for the added-row reconstruction,  and consider different configurations to attach one row of Au atoms on the reconstruction in \Fig{1Au+TiO}. After geometry relaxation, both the Au row in the missing row of Ti and on the surface have large, but similar, energies (a difference of 0.8\,meV/\AA$^2$).  The increase in surface energy is entirely due to the first \TiO\ layer, suggesting that further reduction to \TiOr\  is unfavorable \textit{without} an interfacial layer of gold to ``protect'' the surface.

\begin{figure}[htp]
 \includegraphics[width=3in]{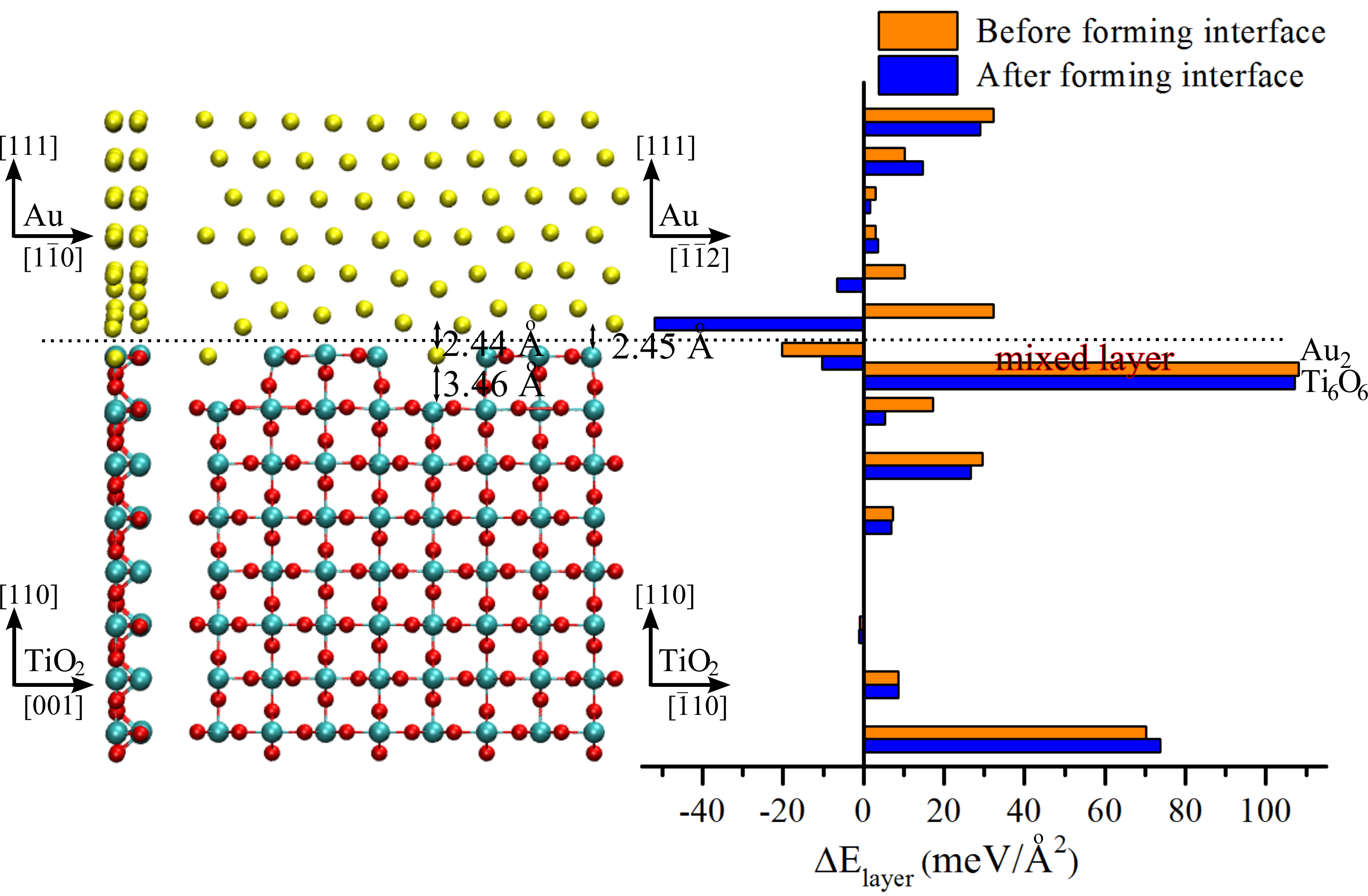}
\caption{Geometry and energy of the Au(111)//\TiOr\ reconstruction following relaxation.  Atomic energy per layer in the reference of bulk value is given before (orange) and after (blue) forming the interface.  The interfacial distance between Au layer with mixed layer is 2.45\,\AA, compared with experimental observation of $2.35\pm0.16\,$\AA.  The work of adhesion is 107\,meV/\AA$^2$ from energy density integration compared with the Au(111) and \TiOr\ reconstruction filled with a Au row.  The stability of the interface comes from a reduction in the Au surface energy with no penalty in the mixed layer, as occurs with the added-row reconstruction.}
\label{fig:TiO}
\end{figure}

\Fig{TiO} shows the geometry of the relaxed Au(111)//\TiOr\ reconstruction interface. Despite the higher energy of the \TiOr\ reconstruction, it produces an \textit{attractive} interface configuration with Au(111).  The interfacial distance between the Au layer and mixed interfacial layer is 2.44--2.45\,\AA; the closer attachment distance compared with the added-row reconstruction is due to the removed oxygen atoms in the interfacial layer.  From total energy, the work of adhesion of the interface is 99\,meV/\AA$^2$.  We integrate the energy density over two Au interfacial layers, one mixed interfacial layer and three next \TiO\ interfacial layers and subtract the corresponding energy density integration in Au layers and the ground-state configuration of an Au row on \TiOr, \Fig{1Au+TiO}(a).  This energy density calculation gives a work of adhesion of $107\pm1\,$meV/\AA$^2$; the difference with the total energy calculation is due to spurious changes in the free \TiO\ surface that the energy density calculation removes. We observe a remarkable drop of atomic energy on Au interfacial layer.  In addition, the mixed layer energy sees only a small change leading to a stabilized interface.  To compute the true work of adhesion, however, we must account for the energy change due to a further reduction from the added-row reconstruction to the \TiOr\ reconstruction.

\begin{figure}[htp]
 \includegraphics[width=3in]{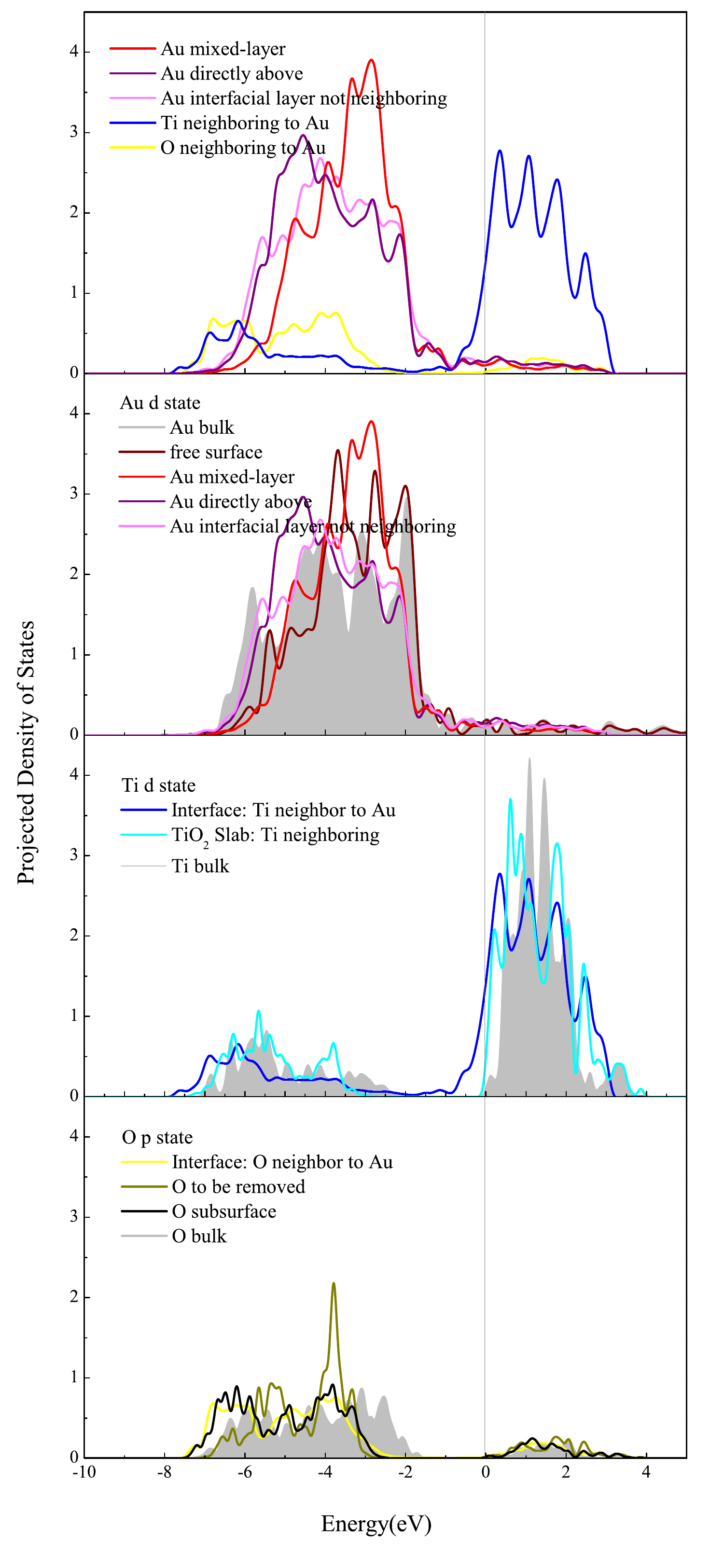}
\caption{Partial electronic density of states for Au, Ti, and O in the relaxed Au(111)//\TiOr\ reconstruction.  The panels show, from top to bottom: total density of states for three types of Au atoms, and Ti and O atoms that neighbor Au in the interface; $d$ density of states for several different Au atomic environments; $d$ density of states for different Ti atomic environments; and $p$ density of states for different O atomic environments.}
\label{fig:pDOS}
\end{figure}

\Fig{pDOS} shows the changes in local electronic density of states for atoms in the Au(111)//\TiOr\ interface compared with other atomic configurations in Au and \TiO.  In the interface, the Au atom mixed in the \TiO\ layer has a narrower width, indicating reduced bonding to neighbors than even Au atoms in the interfacial layer above.  Moreover, the Au $d$ states are pushed towards the Fermi level, even compared with atoms on a free surface.  The widening of the density of states for Au atoms in the interface compared with the free surface corresponds to changes in atomic energy in \Fig{TiO}.  Titanium has a downward shift in unoccupied states pulling them below the Fermi energy in the interface.  Finally, the oxygen atom in the surface next to Au (c.f., \Fig{AddedRow}) that is removed in the new reconstruction sees its density of states narrow and produce a peak; this increase in energy corresponds to the atomic energy changes also seen for this atom.  After removal, the remaining oxygen neighbors have bonding environments that are less disturbed by the presence of Au in the interfacial layer.

\subsubsection{Work of adhesion}
\begin{figure}[htp]
 \includegraphics[width=3in]{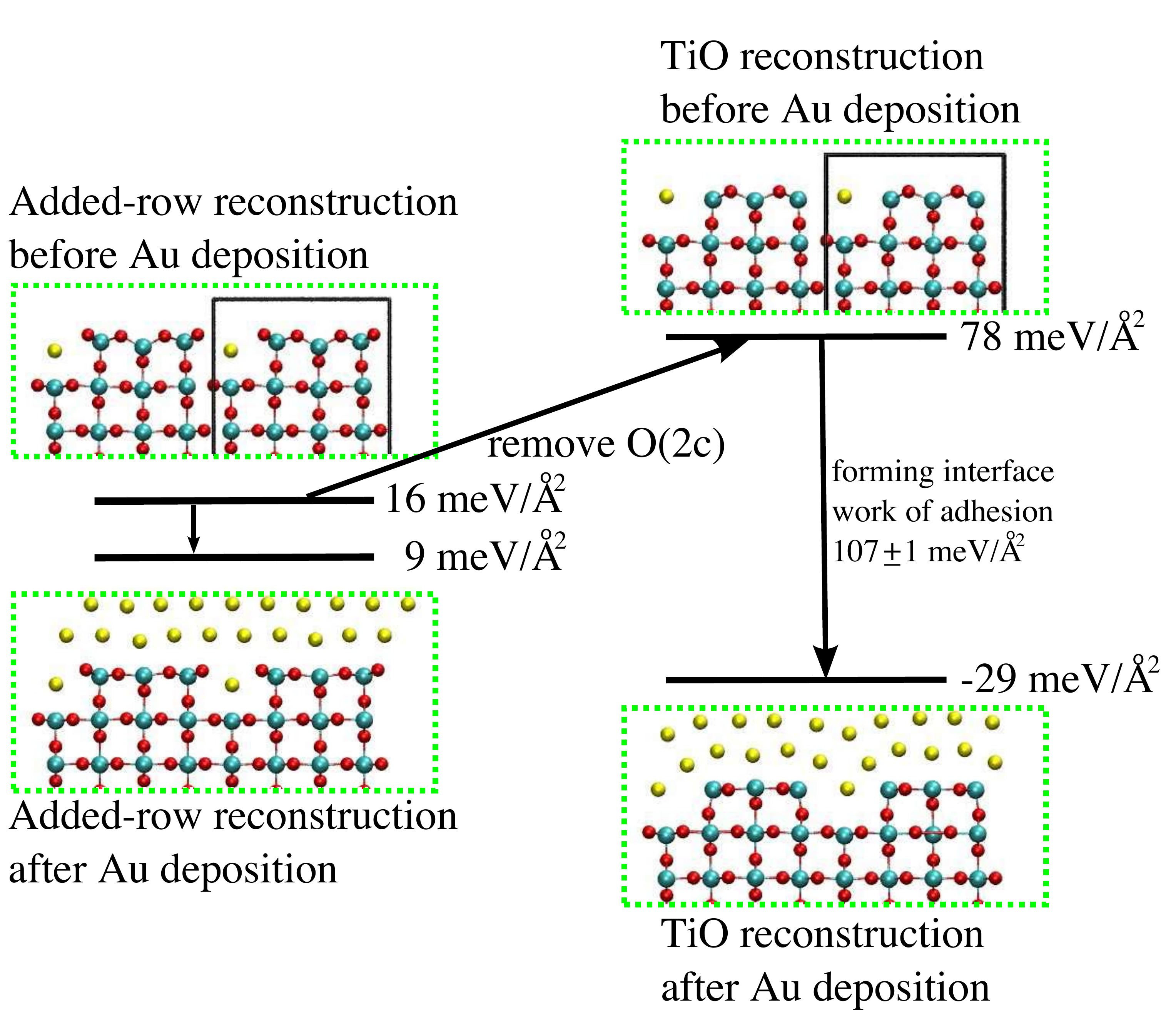}
\caption{Evaluation of work of adhesion for Au(111) on $1\x2$ \TiO (110) reconstructions.  The top two energies are changes in the surface energy before the interface is formed, and are relative to the stable $1\x2$ \TiO (110) added-row reconstruction; hence, we start by adding 16meV/\AA$^2$ when Au is added into the surface (c.f., \Fig{1Au+TiO2AddedRow}).  The bottom two energies are relative to the Au(111) surface and the \TiO (110) surface---the negative work of adhesion.  The \TiOr\ reconstruction leads to a stable interface after Au deposition as the energy required to remove additional oxygen atoms from the added-row reconstruction is offset by a larger reduction in energy when forming the interface.  This is an interesting example of an interfacial reconstruction that is stabilized solely in the presence of the interface.  Compared with the other simple added-row reconstruction which produces a small work of adhesion due to distortions in the mixed layer, the \TiOr\ interfacial reconstruction explains the observed $1\x2$ reconstruction, the interlayer spacing, and is energetically favorable.}
\label{fig:AddedRowVSTiO}
\end{figure}

\Fig{AddedRowVSTiO} shows the relative energies for the different configurations to produce the two different $1\x2$ reconstructions of Au(111)//\TiO (110).  Au(111) adhered to the \TiOr\ reconstruction is the most stable interface configuration with an interfacial distance 2.45\,\AA\ that agrees with the STEM observed\cite{Au-TiO2:Shankar} $2.35\pm0.16\,$\AA.  However, the work of adhesion of 107\,meV/\AA$^2$ is relative to the higher energy \TiOr\ surface with the introduced Au into the subsurface.  The difference between the added-row reconstruction and the \TiOr\ reconstruction means that a single Au row on the \TiOr\ reconstruction is less stable by 62meV/\AA$^2$, plus 16meV/\AA$^2$ to place Au in the subsurface (c.f. \Fig{1Au+TiO2AddedRow}); hence, the \TiOr\ reconstruction produces a stable configuration with work of adhesion of 29\,meV/\AA$^2$ after Au deposition.  Note that we have computed our work of adhesion relative to the stable Au(111) surface with energy 43meV/\AA$^2$ ($38\pm1$meV/\AA$^2$ for the strained surface) and the $1\x2$ added-row reconstruction for \TiO (110) with an energy of $80\pm1$meV/\AA$^2$.  This is lower than simply adhering to the added-row reconstruction, which has a work of adhesion of --9\,meV/\AA$^2$.  It should be noted that the intermediate configuration of \TiOr\ without Au(111) is unstable, and is needed to compute relative energies; given the higher surface energy, it is unlikely that further oxygen reduction occurs before the growth of Au(111) layers.

\subsection{Au(100)//\TiO(110) $1\x 1$: Stoichiometric and reduced interfaces}
Both interfaces on $1\x1$ surfaces use a similar geometry for relaxation.  Along the direction Au$[01\bar1]$//\TiO[001], a single repeat length of Au and \TiO\ gives a 1\%\ lattice mismatch.  Along the direction Au[011]//\TiO$[\bar110]$,  a repeat length of 9 for Au matches with a repeat length of 4 for \TiO, producing a 0.9\%\ lattice mismatch.  The supercell contains 54 Au, 64 Ti, and 128 O atoms for stoichiometric case, and 4 fewer O atoms for reduced case.  As before, we determine the interlayer spacing at the interface following relaxation; with energy density calculations, we can ignore any spurious energy changes due to the opposing Au and \TiO\ surfaces.

\begin{figure}[htp]
 \includegraphics[width=3in]{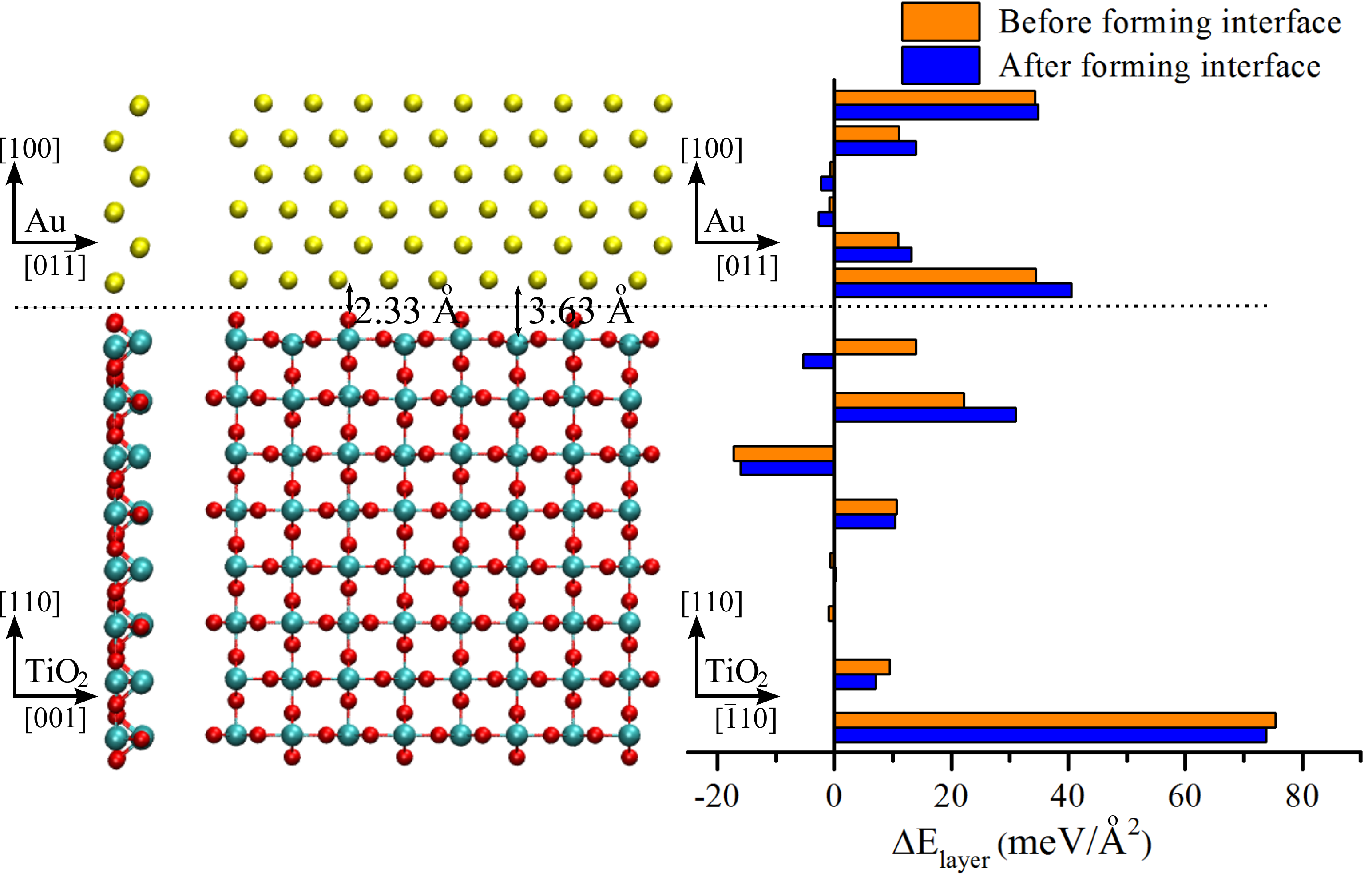}
\caption{Geometry and energy of Au(100) on the stoichiometric \TiO (110) surface following relaxation.  The atomic energy on each layer is referenced to the bulk, and shown before (orange) and after (blue) forming the interface.  The interfacial distance is 3.63\,\AA\ between Au and Ti layers, and 2.33\,\AA\ between Au and bridging O layers.}
\label{fig:100+bO}
\end{figure}

\begin{figure}[htp]
 \includegraphics[width=3in]{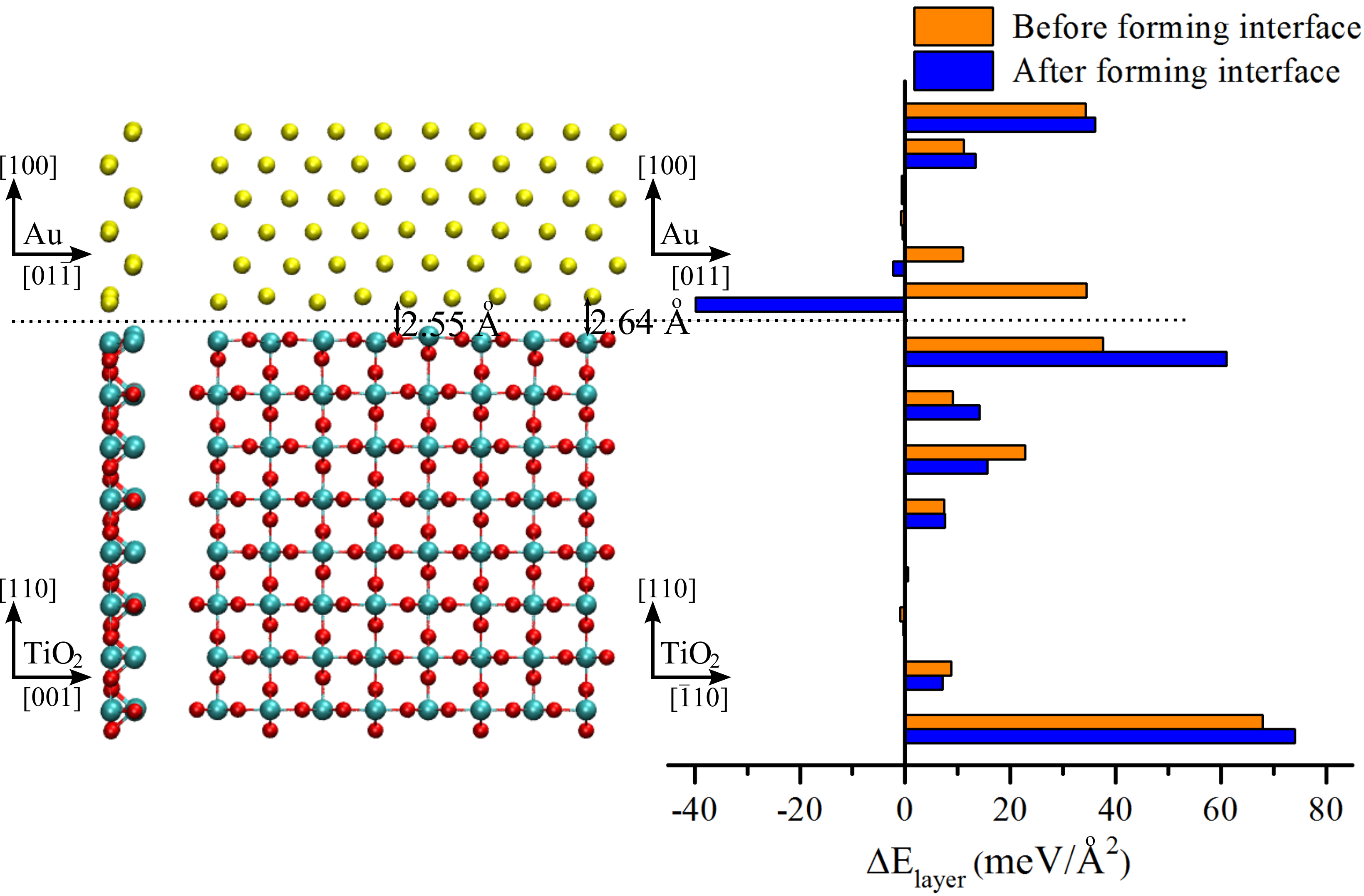}
\caption{Geometry and energy of Au(100) on the reduced \TiO (110) surface following relaxation.  The atomic energy on each layer is referenced to the bulk, and shown before (orange) and after (blue) forming the interface.  The interfacial distance is 2.64\,\AA\ between Au and Ti layers, and 2.55\,\AA\ between Au and in-plane O layers.  The Au layer energy is reduced while the \TiO\ layer energy increases for a work of adhesion of $64\pm1\,$meV/\AA$^2$.}
\label{fig:100-bO}
\end{figure}

\Fig{100+bO} and \Fig{100-bO} show the geometry of the relaxed Au(100) on stoichiometric and reduced \TiO (110) surfaces.  The interfacial distance between Au and Ti layers relaxed to 3.63\,\AA\ with stoichiometric \TiO\ surface, and 2.64\,\AA\ in the configuration with reduced \TiO\ surface.  From total energy, the work of adhesion is 3\,meV/\AA$^2$ of the interface with stoichiometric \TiO, while the work of adhesion of the interface with the reduced \TiO\ surface is 55\,meV/\AA$^2$.  The differences in interlayer spacing and energy is due to the presence or absence of bridging oxygen atoms on the \TiO\ surface.  Energy density shows that \TiO\ layers reach bulk behavior by the fifth layer from interfaces.  We integrate the energy density over two Au layers and four \TiO\ layers to evaluate the work of adhesion strictly from changes in energy near the interface.  This gives a work of adhesion of $1\pm 1\,$meV/\AA$^2$ to the stoichiometric \TiO\ surface, and $64\pm1\,$meV/\AA$^2$ to the reduced \TiO\ surface.  Similar to Au(111)//\TiO(110) reduced interface, atomic energy at the interface decreases in the Au surface, and increases in the \TiO\ surface in the reduced case during forming the interface, to stabilize the structure more than the stoichiometric case.  The energy of \TiO\ free surface away from the interface experiences a spurious energy changes during the interface formation.  Therefore, the integration of energy density over interfacial region reduces the finite-size error, and provides more accurate work of adhesion or interfacial energy.

\section{Conclusions}

\begin{table}[htp]
  \begin{center}
  \begin{tabular}{rcccc}
&&$E_\text{adh}$ [meV/\AA$^2$]&$d_\text{Au-Ti}$ [\AA]&misfit\\
\hline
Stoichiometric&$1\x1$	&$7\mid4\pm1$	&3.90	&3.6\%\\
Reduced&$1\x1$		&$54\mid53\pm1$	&2.79	&3.6\%\\
\hline
Added-row&$1\x2$	&$-9\mid-9\pm1$	&3.00	&2.9\%\\
\TiOr&$1\x2$		&$22\mid49\pm1$	&2.45	&2.9\%\\
\hline
Experiment\cite{Au-TiO2:Shankar,Au-TiO2:Shankar2010}&$1\x2$	&$28\pm7$	&$2.35\pm0.16$\\
\end{tabular}
\end{center}
\caption{Comparison of different Au(111)//\TiO(110) interfaces.  The two different work of adhesions are from the total energy calculation of \Eqn{totalE}, and the energy density integration; the latter accounts for finite-size errors in the supercell calculation.  The Au(111)//\TiOr\ reconstruction agrees with experimental observation in three factors: the $1\x2$ symmetry, the work of adhesion $E_\text{adh}$, and the Au-Ti separation distance $d_\text{Au-Ti}$.  The work of adhesion $29\pm1\,$meV/\AA$^2$ compares well with $28\pm7\,$meV/\AA$^2$; the experimental value\cite{Au-TiO2:Shankar2010} comes from the Wulff-Kaishew theorem\cite{wulff} where $\Delta h/h=E_\text{adh}/\gamma_\text{Au(111)}$, and the geometry parameter $\Delta h/h$ characterizes various equilibrium shapes of supported Au nanoparticles in experiments.  The surface energy $\gamma_\text{Au(111)}$ is 43\,meV/\AA$^2$ from our PAW-GGA-PBE calculation.}
\label{tab:Au111-TiO2}
\end{table}

\Tab{Au111-TiO2} summarizes the geometric and energy comparison of proposed Au$(111)$//\TiO$(110)$ interfaces and the experimental observations\cite{Au-TiO2:Shankar,Au-TiO2:Shankar2010}.  Density functional theory energy density calculations of several Au/\TiO\ interfacial reconstructions determines the equilibrium structure that matches experimental measurements.  Both Au(111) and (100) prefer attaching to reduced rutile \TiO (110) surfaces over stoichiometric surfaces.  Comparison of Au(111) attaching on two \TiO (110) $1\x2$ reconstruction cells shows that the \TiOr\ reconstruction leads to the most stable interface configuration with interfacial distance 2.45\,\AA, and work of adhesion 29\,meV/\AA$^2$.  Atomic energy variation during interface formation demonstrates that the attraction of top Au interfacial layer leads to a stable structure.  The energy density computation also identifies spurious changes to atomic energies on the free-surfaces during the formation of an interface, which affect the computation of work of adhesion from total energy calculations; these finite-size errors are removed.  Our calculations provide an atomistic-level explanation of the stability of the unusual \TiOr\ reconstruction, where further reduction of the interface is possible when ``protected'' by an epitaxial gold layer, and demonstrates the power of energy density computation to guide the identification of stable defect structures.

\begin{acknowledgement}
This research was supported by NSF under grant number DMR-1006077 and through the Materials Computation Center at UIUC, NSF DMR-0325939, and with computational resources from NSF/TeraGrid provided by NCSA and TACC.  We thank J.~M.~Zuo and S.~Sivaramkrishnan for discussion of the experimental results, and R.~M.~Martin for helpful discussions.
\end{acknowledgement}

\providecommand*\mcitethebibliography{\thebibliography}
\csname @ifundefined\endcsname{endmcitethebibliography}
  {\let\endmcitethebibliography\endthebibliography}{}



\begin{mcitethebibliography}{36}
\providecommand*\natexlab[1]{#1}
\providecommand*\mciteSetBstSublistMode[1]{}
\providecommand*\mciteSetBstMaxWidthForm[2]{}
\providecommand*\mciteBstWouldAddEndPuncttrue
  {\def\EndOfBibitem{\unskip.}}
\providecommand*\mciteBstWouldAddEndPunctfalse
  {\let\EndOfBibitem\relax}
\providecommand*\mciteSetBstMidEndSepPunct[3]{}
\providecommand*\mciteSetBstSublistLabelBeginEnd[3]{}
\providecommand*\EndOfBibitem{}
\mciteSetBstSublistMode{f}
\mciteSetBstMaxWidthForm{subitem}{(\alph{mcitesubitemcount})}
\mciteSetBstSublistLabelBeginEnd
  {\mcitemaxwidthsubitemform\space}
  {\relax}
  {\relax}

\bibitem[Haruta et~al.(1987)Haruta, Kobayashi, Sano, and
  Yamada]{Au-TiO2:Haruta1987}
Haruta,~M.; Kobayashi,~T.; Sano,~H.; Yamada,~N. \emph{Chem. Lett.}
  \textbf{1987}, \emph{2}, 405\relax
\mciteBstWouldAddEndPuncttrue
\mciteSetBstMidEndSepPunct{\mcitedefaultmidpunct}
{\mcitedefaultendpunct}{\mcitedefaultseppunct}\relax
\EndOfBibitem
\bibitem[Haruta et~al.(1989)Haruta, Yamada, Kobayashi, and
  Iijima]{Au-TiO2:Haruta1989}
Haruta,~M.; Yamada,~N.; Kobayashi,~T.; Iijima,~S. \emph{J. Catal.}
  \textbf{1989}, \emph{115}, 301\relax
\mciteBstWouldAddEndPuncttrue
\mciteSetBstMidEndSepPunct{\mcitedefaultmidpunct}
{\mcitedefaultendpunct}{\mcitedefaultseppunct}\relax
\EndOfBibitem
\bibitem[Valden et~al.(1998)Valden, Lai, and Goodman]{Au-TiO2:Valden1998}
Valden,~M.; Lai,~X.; Goodman,~D.~W. \emph{Science} \textbf{1998}, \emph{281},
  1647\relax
\mciteBstWouldAddEndPuncttrue
\mciteSetBstMidEndSepPunct{\mcitedefaultmidpunct}
{\mcitedefaultendpunct}{\mcitedefaultseppunct}\relax
\EndOfBibitem
\bibitem[Boccuzzi et~al.(1999)Boccuzzi, Chiorino, Manzoli, Andreeva, and
  Tabakova]{Au-TiO2:Boccuzzi1989}
Boccuzzi,~F.; Chiorino,~A.; Manzoli,~M.; Andreeva,~D.; Tabakova,~T. \emph{J.
  Catal.} \textbf{1999}, \emph{188}, 176\relax
\mciteBstWouldAddEndPuncttrue
\mciteSetBstMidEndSepPunct{\mcitedefaultmidpunct}
{\mcitedefaultendpunct}{\mcitedefaultseppunct}\relax
\EndOfBibitem
\bibitem[Hayashi et~al.(1998)Hayashi, Tanaka, and Haruta]{Au-TiO2:Hayashi1998}
Hayashi,~T.; Tanaka,~K.; Haruta,~M. \emph{J. Catal.} \textbf{1998}, \emph{178},
  566\relax
\mciteBstWouldAddEndPuncttrue
\mciteSetBstMidEndSepPunct{\mcitedefaultmidpunct}
{\mcitedefaultendpunct}{\mcitedefaultseppunct}\relax
\EndOfBibitem
\bibitem[Diebold(2003)]{TiO2:SSReport}
Diebold,~U. \emph{Surf. Sci. Rep.} \textbf{2003}, \emph{48}, 53\relax
\mciteBstWouldAddEndPuncttrue
\mciteSetBstMidEndSepPunct{\mcitedefaultmidpunct}
{\mcitedefaultendpunct}{\mcitedefaultseppunct}\relax
\EndOfBibitem
\bibitem[Bond and Thompson(2000)Bond, and Thompson]{Au-TiO2:Bond2000}
Bond,~G.~C.; Thompson,~D.~T. \emph{Gold Bull.} \textbf{2000}, \emph{33},
  41\relax
\mciteBstWouldAddEndPuncttrue
\mciteSetBstMidEndSepPunct{\mcitedefaultmidpunct}
{\mcitedefaultendpunct}{\mcitedefaultseppunct}\relax
\EndOfBibitem
\bibitem[Grunwaldt et~al.(1999)Grunwaldt, Maciejewski, Becker, Fabrizioli, and
  Baiker]{Au-TiO2:Grunwaldt1999}
Grunwaldt,~J.-D.; Maciejewski,~M.; Becker,~O.; Fabrizioli,~P.; Baiker,~A.
  \emph{J. Catal.} \textbf{1999}, \emph{186}, 458\relax
\mciteBstWouldAddEndPuncttrue
\mciteSetBstMidEndSepPunct{\mcitedefaultmidpunct}
{\mcitedefaultendpunct}{\mcitedefaultseppunct}\relax
\EndOfBibitem
\bibitem[Haruta(2004)]{Au-TiO2:Haruta2004}
Haruta,~M. \emph{Gold Bull.} \textbf{2004}, \emph{37}, 27\relax
\mciteBstWouldAddEndPuncttrue
\mciteSetBstMidEndSepPunct{\mcitedefaultmidpunct}
{\mcitedefaultendpunct}{\mcitedefaultseppunct}\relax
\EndOfBibitem
\bibitem[Cosandey and Madey(2001)Cosandey, and Madey]{Au-TiO2:Cosandey2001}
Cosandey,~F.; Madey,~T.~E. \emph{Surf. Rev. Lett.} \textbf{2001}, \emph{8},
  73\relax
\mciteBstWouldAddEndPuncttrue
\mciteSetBstMidEndSepPunct{\mcitedefaultmidpunct}
{\mcitedefaultendpunct}{\mcitedefaultseppunct}\relax
\EndOfBibitem
\bibitem[Sivaramakrishnan et~al.(2011)Sivaramakrishnan, Yu, Pierce, Scarpelli,
  Wen, Trinkle, and Zuo]{Au-TiO2:Shankar}
Sivaramakrishnan,~S.; Yu,~M.; Pierce,~B.~J.; Scarpelli,~M.~E.; Wen,~J.;
  Trinkle,~D.~R.; Zuo,~J.-M. (under review, Science)\relax
\mciteBstWouldAddEndPuncttrue
\mciteSetBstMidEndSepPunct{\mcitedefaultmidpunct}
{\mcitedefaultendpunct}{\mcitedefaultseppunct}\relax
\EndOfBibitem
\bibitem[Akita et~al.(2008)Akita, Tanaka, Kohyama, and
  Haruta]{Au-TiO2:Akita2008}
Akita,~T.; Tanaka,~K.; Kohyama,~M.; Haruta,~M. \emph{Surf. Interface Anal.}
  \textbf{2008}, \emph{40}, 1760\relax
\mciteBstWouldAddEndPuncttrue
\mciteSetBstMidEndSepPunct{\mcitedefaultmidpunct}
{\mcitedefaultendpunct}{\mcitedefaultseppunct}\relax
\EndOfBibitem
\bibitem[Kohn and Sham(1965)Kohn, and Sham]{Theor:KS}
Kohn,~W.; Sham,~L.~J. \emph{Phys. Rev.} \textbf{1965}, \emph{140}, A1133\relax
\mciteBstWouldAddEndPuncttrue
\mciteSetBstMidEndSepPunct{\mcitedefaultmidpunct}
{\mcitedefaultendpunct}{\mcitedefaultseppunct}\relax
\EndOfBibitem
\bibitem[Yang et~al.(2000)Yang, Wu, and Goodman]{Au-TiO2:Yang2000}
Yang,~Z.; Wu,~R.; Goodman,~D.~W. \emph{Phys. Rev. B} \textbf{2000}, \emph{61},
  14066\relax
\mciteBstWouldAddEndPuncttrue
\mciteSetBstMidEndSepPunct{\mcitedefaultmidpunct}
{\mcitedefaultendpunct}{\mcitedefaultseppunct}\relax
\EndOfBibitem
\bibitem[Wang and Hwang(2003)Wang, and Hwang]{Au-TiO2:Wang2003}
Wang,~Y.; Hwang,~G.~S. \emph{Surf. Sci.} \textbf{2003}, \emph{542}, 72\relax
\mciteBstWouldAddEndPuncttrue
\mciteSetBstMidEndSepPunct{\mcitedefaultmidpunct}
{\mcitedefaultendpunct}{\mcitedefaultseppunct}\relax
\EndOfBibitem
\bibitem[Amrendra et~al.(2003)Amrendra, Greg, and Horia]{Au-TiO2:Amrendra2003}
Amrendra,~V.; Greg,~M.; Horia,~M. \emph{J. Chem. Phys.} \textbf{2003},
  \emph{118}, 6536\relax
\mciteBstWouldAddEndPuncttrue
\mciteSetBstMidEndSepPunct{\mcitedefaultmidpunct}
{\mcitedefaultendpunct}{\mcitedefaultseppunct}\relax
\EndOfBibitem
\bibitem[Wahlstr\"{o}m et~al.(2003)Wahlstr\"{o}m, Lopez, Schaub, Thostrup,
  R\o{}nnau, Africh, L\ae{}gsgaard, N\o{}rskov, and
  Besenbacher]{Au-TiO2:Wahlstrom2003}
Wahlstr\"{o}m,~E.; Lopez,~N.; Schaub,~R.; Thostrup,~P.; R\o{}nnau,~A.;
  Africh,~C.; L\ae{}gsgaard,~E.; N\o{}rskov,~J.~K.; Besenbacher,~F. \emph{Phys.
  Rev. Lett.} \textbf{2003}, \emph{90}, 026101\relax
\mciteBstWouldAddEndPuncttrue
\mciteSetBstMidEndSepPunct{\mcitedefaultmidpunct}
{\mcitedefaultendpunct}{\mcitedefaultseppunct}\relax
\EndOfBibitem
\bibitem[Okazaki et~al.(2004)Okazaki, Morikawa, Tanaka, Tanaka, and
  Kohyama]{Au-TiO2:Okazaki2004}
Okazaki,~K.; Morikawa,~Y.; Tanaka,~S.; Tanaka,~K.; Kohyama,~M. \emph{Phys. Rev.
  B} \textbf{2004}, \emph{69}, 235404\relax
\mciteBstWouldAddEndPuncttrue
\mciteSetBstMidEndSepPunct{\mcitedefaultmidpunct}
{\mcitedefaultendpunct}{\mcitedefaultseppunct}\relax
\EndOfBibitem
\bibitem[Lopez et~al.(2004)Lopez, N\o{}rskov, Janssens, Carlsson, Puig-Molina,
  Clausen, and Grunwaldt]{Au-TiO2:Lopez2004}
Lopez,~N.; N\o{}rskov,~J.~K.; Janssens,~T. V.~W.; Carlsson,~A.;
  Puig-Molina,~A.; Clausen,~B.~S.; Grunwaldt,~J.-D. \emph{J. Catal.}
  \textbf{2004}, \emph{225}, 86\relax
\mciteBstWouldAddEndPuncttrue
\mciteSetBstMidEndSepPunct{\mcitedefaultmidpunct}
{\mcitedefaultendpunct}{\mcitedefaultseppunct}\relax
\EndOfBibitem
\bibitem[Pabisiak and Kiejna(2009)Pabisiak, and Kiejna]{Au-TiO2:Pabisiak2009}
Pabisiak,~T.; Kiejna,~A. \emph{Phys. Rev. B} \textbf{2009}, \emph{79},
  085411\relax
\mciteBstWouldAddEndPuncttrue
\mciteSetBstMidEndSepPunct{\mcitedefaultmidpunct}
{\mcitedefaultendpunct}{\mcitedefaultseppunct}\relax
\EndOfBibitem
\bibitem[Pillay and Hwang(2005)Pillay, and Hwang]{Au-TiO2:Pillay2005}
Pillay,~D.; Hwang,~G.~S. \emph{Phys. Rev. B} \textbf{2005}, \emph{72},
  205422\relax
\mciteBstWouldAddEndPuncttrue
\mciteSetBstMidEndSepPunct{\mcitedefaultmidpunct}
{\mcitedefaultendpunct}{\mcitedefaultseppunct}\relax
\EndOfBibitem
\bibitem[Shi et~al.(2009)Shi, Kohyama, Tanaka, and Takeda]{Au-TiO2:Shi2009}
Shi,~H.; Kohyama,~M.; Tanaka,~S.; Takeda,~S. \emph{Phys. Rev. B} \textbf{2009},
  \emph{80}, 155413\relax
\mciteBstWouldAddEndPuncttrue
\mciteSetBstMidEndSepPunct{\mcitedefaultmidpunct}
{\mcitedefaultendpunct}{\mcitedefaultseppunct}\relax
\EndOfBibitem
\bibitem[Shibata et~al.(2009)Shibata, Goto, Matsunaga, Mizoguchi, Findlay,
  Yamamoto, and Ikuhara]{Au-TiO2:Shibata2009}
Shibata,~N.; Goto,~A.; Matsunaga,~K.; Mizoguchi,~T.; Findlay,~S.~D.;
  Yamamoto,~T.; Ikuhara,~Y. \emph{Phys. Rev. Lett.} \textbf{2009}, \emph{102},
  136105\relax
\mciteBstWouldAddEndPuncttrue
\mciteSetBstMidEndSepPunct{\mcitedefaultmidpunct}
{\mcitedefaultendpunct}{\mcitedefaultseppunct}\relax
\EndOfBibitem
\bibitem[Pang et~al.(1998)Pang, Haycock, Raza, Murray, Thornton, G\"ulseren,
  James, and Bullett]{TiO2:AddedRow}
Pang,~C.~L.; Haycock,~S.~A.; Raza,~H.; Murray,~P.~W.; Thornton,~G.;
  G\"ulseren,~O.; James,~R.; Bullett,~D.~W. \emph{Phys. Rev. B} \textbf{1998},
  \emph{58}, 1586\relax
\mciteBstWouldAddEndPuncttrue
\mciteSetBstMidEndSepPunct{\mcitedefaultmidpunct}
{\mcitedefaultendpunct}{\mcitedefaultseppunct}\relax
\EndOfBibitem
\bibitem[Chetty and Martin(1992)Chetty, and Martin]{EDM:ChettyMartin}
Chetty,~N.; Martin,~R.~M. \emph{Phys. Rev. B} \textbf{1992}, \emph{45},
  6074\relax
\mciteBstWouldAddEndPuncttrue
\mciteSetBstMidEndSepPunct{\mcitedefaultmidpunct}
{\mcitedefaultendpunct}{\mcitedefaultseppunct}\relax
\EndOfBibitem
\bibitem[Yu et~al.(2011)Yu, Trinkle, and Martin]{EDM:PAW}
Yu,~M.; Trinkle,~D.~R.; Martin,~R.~M. \emph{Phys. Rev. B} \textbf{2011},
  (accepted; arXiv:1011.4683)\relax
\mciteBstWouldAddEndPuncttrue
\mciteSetBstMidEndSepPunct{\mcitedefaultmidpunct}
{\mcitedefaultendpunct}{\mcitedefaultseppunct}\relax
\EndOfBibitem
\bibitem[Sivaramakrishnan et~al.(2010)Sivaramakrishnan, Wen, Scarpelli, Pierce,
  and Zuo]{Au-TiO2:Shankar2010}
Sivaramakrishnan,~S.; Wen,~J.; Scarpelli,~M.~E.; Pierce,~B.~J.; Zuo,~J.-M.
  \emph{Phys. Rev. B} \textbf{2010}, \emph{82}, 195421\relax
\mciteBstWouldAddEndPuncttrue
\mciteSetBstMidEndSepPunct{\mcitedefaultmidpunct}
{\mcitedefaultendpunct}{\mcitedefaultseppunct}\relax
\EndOfBibitem
\bibitem[Bl\"{o}chl(1994)]{Theor:PAW}
Bl\"{o}chl,~P.~E. \emph{Phys. Rev. B} \textbf{1994}, \emph{50}, 17953\relax
\mciteBstWouldAddEndPuncttrue
\mciteSetBstMidEndSepPunct{\mcitedefaultmidpunct}
{\mcitedefaultendpunct}{\mcitedefaultseppunct}\relax
\EndOfBibitem
\bibitem[Kresse and Furthm\"{u}ller(1996)Kresse, and
  Furthm\"{u}ller]{Theor:VASP}
Kresse,~G.; Furthm\"{u}ller,~J. \emph{Phys. Rev. B} \textbf{1996}, \emph{54},
  11169\relax
\mciteBstWouldAddEndPuncttrue
\mciteSetBstMidEndSepPunct{\mcitedefaultmidpunct}
{\mcitedefaultendpunct}{\mcitedefaultseppunct}\relax
\EndOfBibitem
\bibitem[Kresse and Joubert(1999)Kresse, and Joubert]{Theor:VASP-USPP-PAW}
Kresse,~G.; Joubert,~D. \emph{Phys. Rev. B} \textbf{1999}, \emph{59},
  1758\relax
\mciteBstWouldAddEndPuncttrue
\mciteSetBstMidEndSepPunct{\mcitedefaultmidpunct}
{\mcitedefaultendpunct}{\mcitedefaultseppunct}\relax
\EndOfBibitem
\bibitem[Perdew et~al.(1996)Perdew, Burke, and Ernzerhof]{Theor:PBE}
Perdew,~J.~P.; Burke,~K.; Ernzerhof,~M. \emph{Phys. Rev. Lett.} \textbf{1996},
  \emph{77}, 3865\relax
\mciteBstWouldAddEndPuncttrue
\mciteSetBstMidEndSepPunct{\mcitedefaultmidpunct}
{\mcitedefaultendpunct}{\mcitedefaultseppunct}\relax
\EndOfBibitem
\bibitem[Monkhorst and Pack(1976)Monkhorst, and Pack]{Theor:MP}
Monkhorst,~H.~J.; Pack,~J.~D. \emph{Phys. Rev. B} \textbf{1976}, \emph{13},
  5188\relax
\mciteBstWouldAddEndPuncttrue
\mciteSetBstMidEndSepPunct{\mcitedefaultmidpunct}
{\mcitedefaultendpunct}{\mcitedefaultseppunct}\relax
\EndOfBibitem
\bibitem[Methfessel and Paxton(1989)Methfessel, and Paxton]{Methfessel1989}
Methfessel,~M.; Paxton,~A.~T. \emph{Phys. Rev. B} \textbf{1989}, \emph{40},
  3616\relax
\mciteBstWouldAddEndPuncttrue
\mciteSetBstMidEndSepPunct{\mcitedefaultmidpunct}
{\mcitedefaultendpunct}{\mcitedefaultseppunct}\relax
\EndOfBibitem
\bibitem[Yu and Trinkle(2011)Yu, and Trinkle]{WeightMethod}
Yu,~M.; Trinkle,~D.~R. \emph{J. Chem. Phys.} \textbf{2011}, \emph{134},
  064111\relax
\mciteBstWouldAddEndPuncttrue
\mciteSetBstMidEndSepPunct{\mcitedefaultmidpunct}
{\mcitedefaultendpunct}{\mcitedefaultseppunct}\relax
\EndOfBibitem
\bibitem[Winterbottom(1967)]{wulff}
Winterbottom,~W.~L. \emph{Acta Metall.} \textbf{1967}, \emph{15}, 303\relax
\mciteBstWouldAddEndPuncttrue
\mciteSetBstMidEndSepPunct{\mcitedefaultmidpunct}
{\mcitedefaultendpunct}{\mcitedefaultseppunct}\relax
\EndOfBibitem
\end{mcitethebibliography}
\end{document}